\documentclass{article}
\usepackage[utf8]{inputenc}
\usepackage[english]{babel}
\usepackage{url}
\usepackage{amssymb}
\usepackage{amsmath}
\usepackage{graphicx}
\usepackage{subcaption}

\usepackage[]{algorithm2e}
\usepackage[a4paper,margin=1.5in,footskip=0.25in]{geometry}
\usepackage{enumitem}
\title{\bf Game of Life on Graphs}

\author{%
Mikhail Krechetov\\ 
Skolkovo Institute of Science and Technology,\\
Moscow, Russia, 121205\\
 \texttt{Mikhail.Krechetov@skoltech.ru}
}
 \date{}

\begin{document}

\maketitle

\begin{abstract}
  We consider a specific graph dynamical system inspired by the famous Conway's Game of Life in this work. We study the properties of the dynamical system on different graphs and introduce a new efficient heuristic for graph isomorphism testing. We use the evolution of our system to extract features from a graph in a deterministic way and observe that the extracted features are unique for all connected graphs with up to ten vertices.
\end{abstract}

\section{Introduction}

In this paper, we study the following discrete dynamical system on a graph $G = (V, E)$:
\begin{enumerate}[leftmargin=*]
    \item Vertices of a graph can be in one of the two states: 'alive' or 'dead'.
    \item Initially, some vertices are alive (usually we start with a single 'alive' vertex).
    \item In the next step, any 'alive' vertex with fewer than $\mathfrak a$ 'alive' neighbors dies, as if by underpopulation.
    \item Any 'alive' vertex with less than $\mathfrak d$ 'dead' neighbors dies, as if by overpopulation.
    \item Any 'dead' vertex with exactly $\mathfrak r$ 'alive' neighbors becomes 'alive', as if by reproduction.
    \item We repeat this dynamic for a fixed number of steps or may wait until the population dies or repeats itself.
\end{enumerate}

We call the dynamics above "Game of Life on Graphs" due to its resemblance to the well-known Conway's Game of Life \cite{Life}. Conway's Game of Life is a cellular automaton on an infinite two-dimensional orthogonal grid, and the following rules govern the dynamic:

\begin{enumerate}[leftmargin=*]
    \item Any 'alive' cell with fewer than two live neighbors dies, as if by underpopulation.
    \item Any 'alive' cell with two or three live neighbors lives on to the next generation.
    \item Any 'alive' cell with more than three live neighbors dies, as if by overpopulation.
    \item Any 'dead' cell with exactly three live neighbors becomes a live cell, as if by reproduction.
\end{enumerate}

Conway's Game of Life is Turing complete \cite{Turing2, Turing1} and people have discovered numerous complex patterns of the dynamics above, see for example \url{https://conwaylife.appspot.com/library}. Since it was introduced, the Game of Life has appeared in various contexts: as a two-player game \cite{TwoPlayer}, from the fuzzy logic perspective \cite{Fuzzy}, in quantum annealing simulations \cite{Quantum} and even in learning neural networks \cite{NNHard}.\\

Similar to the Game of Life, according to our dynamics, the vertex becomes alive by reproduction if it has at least one neighbor alive and dies by overpopulation if all its neighbors are alive; we illustrate the definition in Section~\ref{sec:Life}. However, this is not the only resemblance. We observe that our dynamics show very complex behavior on different graphs, and the 'alive' patterns evolution differs significantly from one graph to another. In Section~\ref{sec:GI} we show how the observed 'alive' patterns can be used to build an efficient heuristic for graph isomorphism testing. We discuss the numerical experiments in Section~\ref{sec:exp}.

\section{Life on Graphs}
\label{sec:Life}

This section provides some definitions, examples, and intuition behind the Game of Life on Graphs. In the most general case, we define the dynamics by the four parameters: graph $G = (V, E)$, the sets $A_0$ and $D_0$ of initially alive and initially dead vertices (we will omit $D_0$ in the future since $A_0 \cup D_0 = V$), integer numbers $a$ and $d$ (the meaning of this parameters as defined below).\\

By $x_v(t) \in \{'alive', 'dead'\}$ we define the state of a vertex $v \in V$ at time $t \in \mathbb Z_+$. By $A_t = \{v\in V | x_v(t) = 'alive'\}$ we define the set of all alive vertices at the step $t$. Similarly, by $D_t = \{v\in V | x_v(t) = 'dead'\}$ we define the set of all dead vertices at the step $t$. By $N_G(v) = \{u\in V\setminus v | (u, v) \in E \}$ we denote the set on heighbors of a vertex $v$. Then the Game of Life on Graphs is denoted $GLG(G, A_0, \mathfrak a, \mathfrak d, \mathfrak r)$ and is defined by the following dynamics:

\begin{equation}
x_v(t)=\begin{cases}
			'alive', & \text{if $x_v(t) ='alive'$ and $|A_{t-1} \cap N_G(v)| \ge \mathfrak a$ and $|D_{t-1}\cap N_G(v)| \ge \mathfrak d$}\\
			'alive', & \text{if $x_v(t) ='dead'$ and $|A_{t-1} \cap N_G(v)| == \mathfrak r$}\\
            'dead', & \text{otherwise}
		 \end{cases}\tag{$GLG(G, A_0, \mathfrak a, \mathfrak d, \mathfrak r)$}
\end{equation}

In other words, the dynamical system on the graph evolves in time, and the vertex continue to live iff it has at least $\mathfrak a$ alive adjacent vertices and at least $\mathfrak d$ dead adjacent vertices; if a 'dead' vertex has $mathfrak r$ 'alive' neighbors, it becomes 'alive' by reproduction.\\

{\bf Definition.} We say, that for $GLG(G, A_0, \mathfrak a, \mathfrak d, \mathfrak r)$ the set $A_0$ is the {\it initial population}. For every $t$ we call a set $\{A_v(t) | v\in V\}$ {\it life pattern}.\\

In this work, we mainly work with $\mathfrak a = 1$, $\mathfrak d = 1$ and $\mathfrak r = 1$. We do so only because these parameters were the most successful in testing graph isomorphism, see Section~\ref{sec:GI}. We also fix $|A_0|=1$. It means that we only consider the initial populations of size one; this is enough for all the results reported in this paper. However, one can consider larger initial populations.\\

{\bf Definition.} We say, that for $GLG(G, A_0, \mathfrak a, \mathfrak d, \mathfrak r)$ the initial population $A_0$ {\it dies} (and the Game {\it halts}) if after the finite number of steps $t_d$, the state $x_v(t_d) = 'dead'$ of every vertex $v\in V$.\\

{\bf Definition.} We say, that for $GLG(G, A_0, \mathfrak a, \mathfrak d, \mathfrak r)$ the initial population $A_0$ {\it repeats} if there exists $t_1$ and $t_2$ ($t_1\ne t_2$), so that $x_v(t_1) = x_v(t_2)$ for every $v\in V$. In other words, the evolution repeats itself in cycles and never dies.\\

{\bf Proposition.} For a finite graph, there is a finite number of possible life patterns; thus, every instance of Game of Life on finite graphs either dies or repeats.\\

{\bf Definition.} For a finite graph $G$ and an instance of Game of Life $GLG(G, A_0, \mathfrak a, \mathfrak d, \mathfrak r)$ we count the total number of distinct observed life patterns before the initial population dies or repeats. That number is called the {\it complexity} of the $GLG(G, A_0, a, d)$.\\

\subsection{Examples}
\begin{figure}[ht!]
\centering
\includegraphics[width=0.25\textwidth]{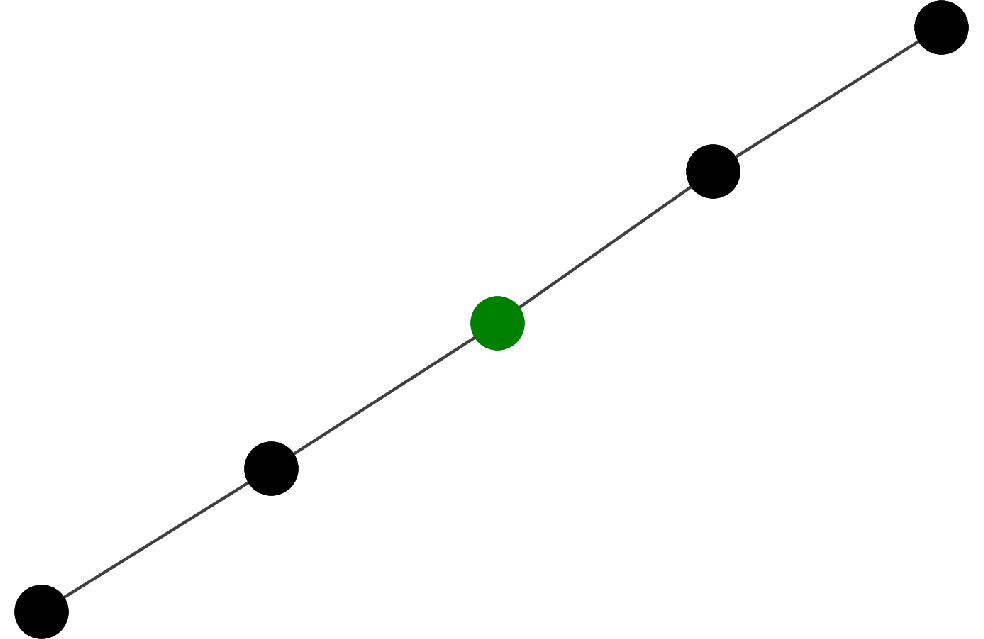}\hfill
\includegraphics[width=0.25\textwidth]{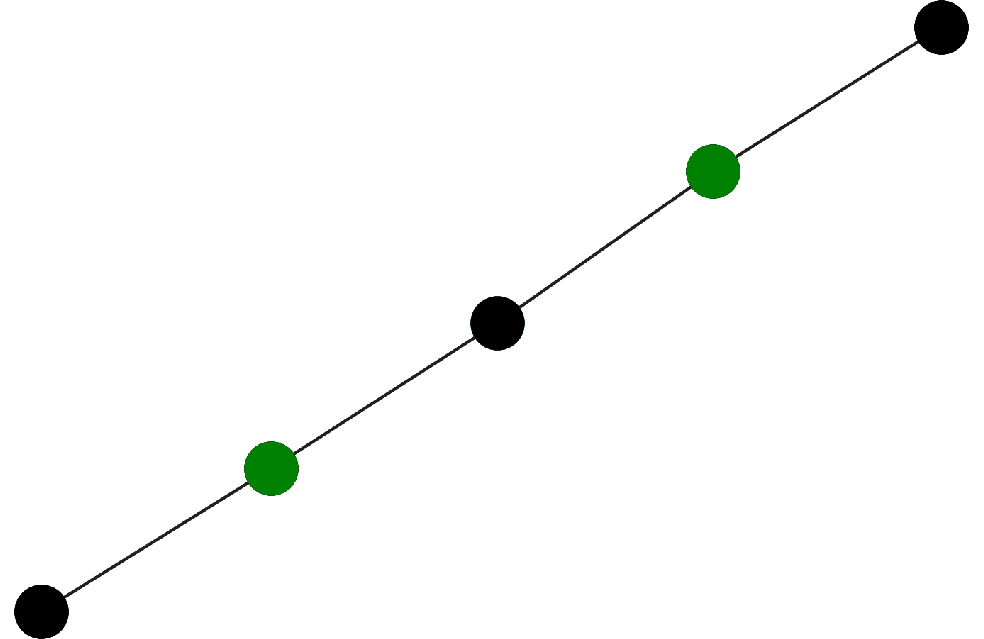}\hfill
\includegraphics[width=0.25\textwidth]{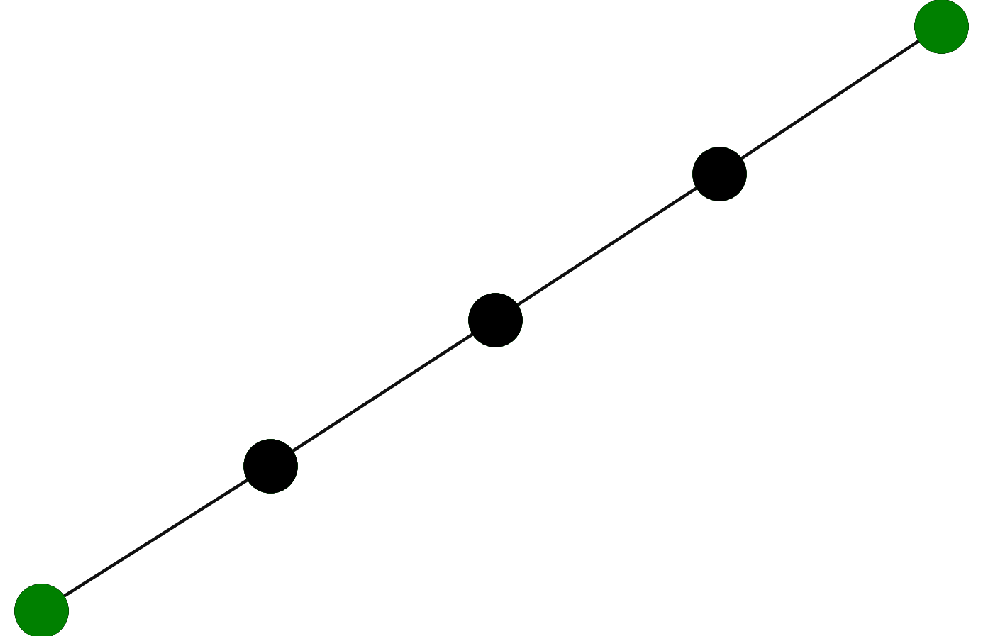}\hfill
\includegraphics[width=0.25\textwidth]{pict/line/l_01.png}
\caption{An exemplary Game of Life on the line graph with five vertices. We start with the initial pattern on the left and proceed for three steps until the game {\it repeats} at the third step. Thus the complexity of this game equals three.}
\label{fig:line0}
\end{figure}

\begin{figure}[ht!]
\centering
\includegraphics[width=0.33\textwidth]{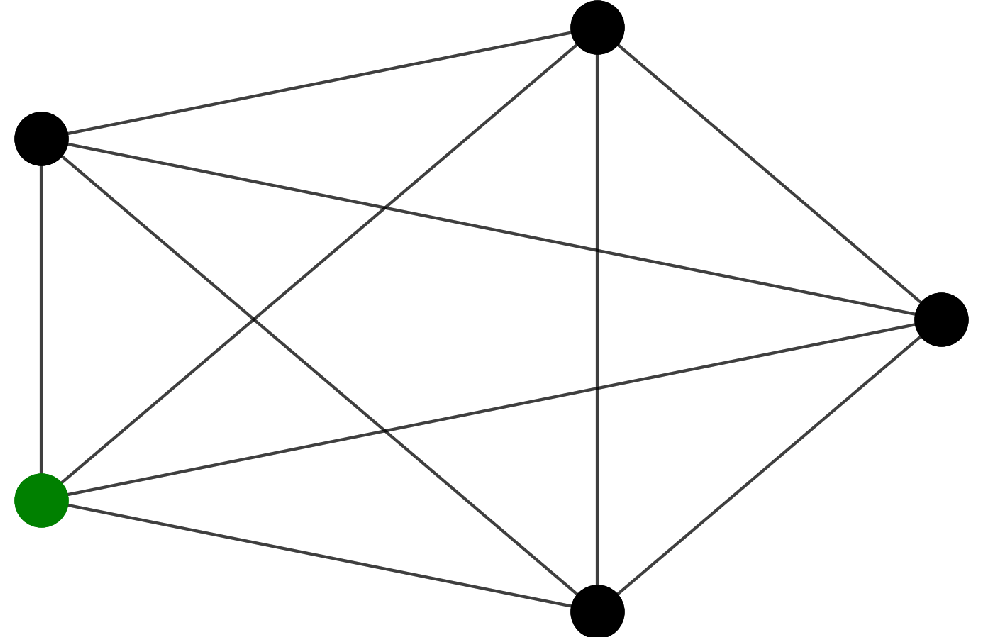}\hfill
\includegraphics[width=0.33\textwidth]{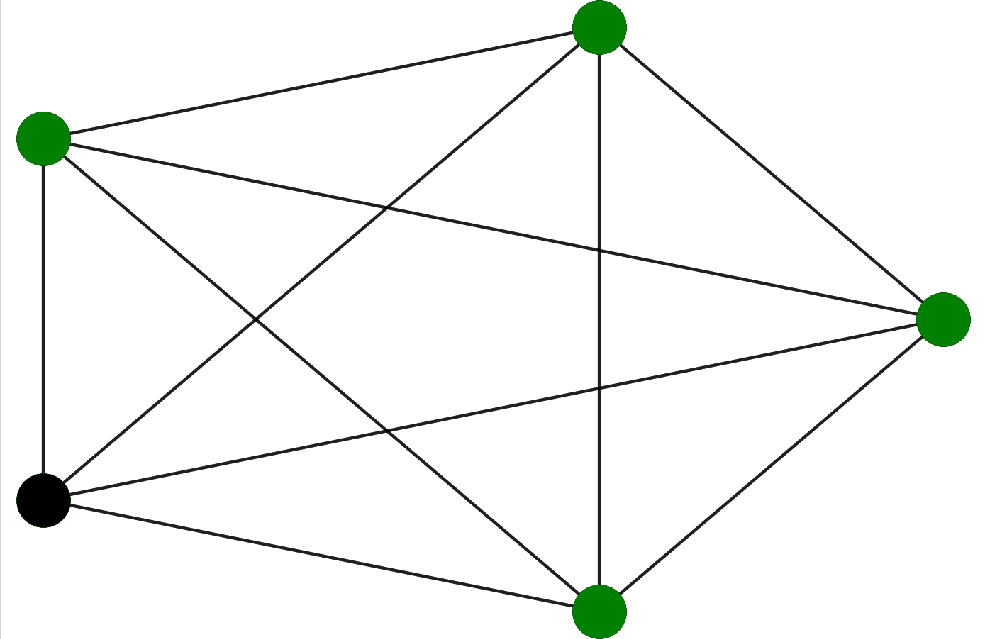}\hfill
\includegraphics[width=0.33\textwidth]{pict/complete/c1.png}
\caption{An exemplary Game of Life on the complete graph with five vertices. We start with the initial pattern on the left and proceed for three steps until the life pattern from the third step repeats the life pattern from the second step. Thus the complexity of this game equals two.}
\label{fig:complete}
\end{figure}

In this section we illustrate all the definitions given above and run \[GLG(G, \{i\}, 1, 1, 1)\] for different graphs $G = (V, E)$ and different initial life patterns with single vertex $i\in V$. In Figures~\ref{fig:line0},\ref{fig:complete} we color 'alive' vertices in green and 'dead' vertices in black.\\

In Figure~\ref{fig:line0} we illustrate the Game of Life on a line graph. In Figure~\ref{fig:complete} we show that a single initially alive node in the complete graph makes all other nodes alive and dies itself; the resulting {\it life pattern} continues to exist forever without changes. More illustrations are available at \url{https://github.com/mkrechetov/GameOfLifeOnGraphs}.

\subsection{Universality}

\begin{figure}[ht!]
\centering
\includegraphics[width=\textwidth]{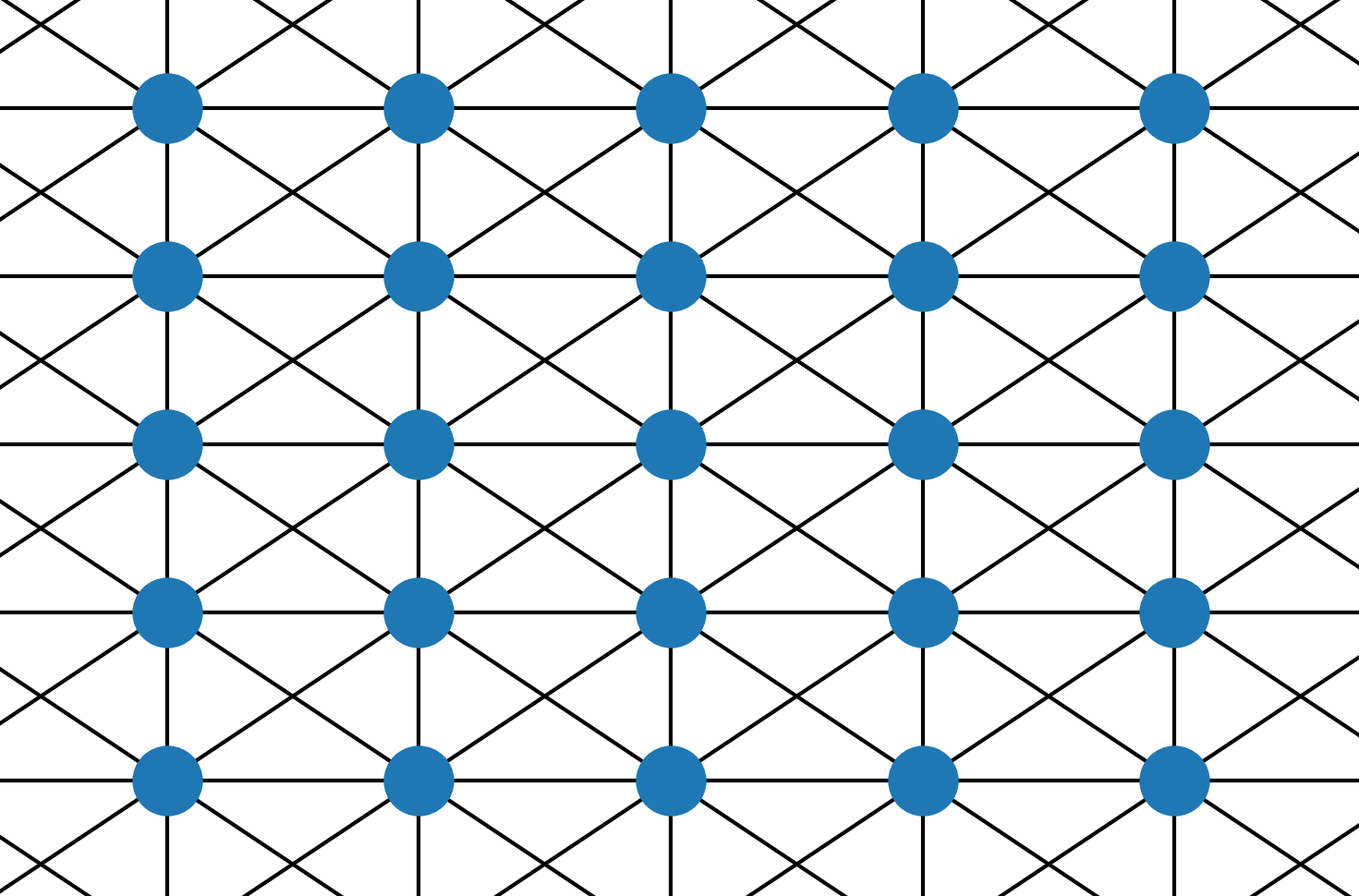}
\caption{An illustration of the infinite grid graph from the Conway's Game of Life.}
\label{fig:GOLgrid}
\end{figure}

Note that Conway's Game of Life is an instance of the Game of Life on Graphs. Let us consider an infinite grid graph from the Figure~\ref{fig:GOLgrid}, a set of initially alive vertices $A_0$ and an instance $GLG(G, A_0, 2, 5, 3)$. This instance of the Game of Life on Graphs is equivalent to Conway's Game of Life; now, the proposition below follows:\\

{\bf Proposition.} In the most general formulation, the Game of Life on Graphs, \[GLG(G, A_0, \mathfrak a, \mathfrak d, \mathfrak r)\] is Turing Complete. 

\subsection{Halting problem}

{\bf Definition.} Let $GLG(G, A_0, \mathfrak a, \mathfrak d, \mathfrak r)$ be an instance of the Game of Life on Graphs. Does the initial population die after some number of iterations or continue to exist forever? We call this question {\it halting} of the Game of Life.\\

Clearly, in the most general formulation (and due to the proposition from the previous section), the {\it halting} of the Game of Life is undecidable. However, for finite graphs, the halting problem is in {\bf EXPTIME} since there is a finite number of possible {\it life patterns}; it means that in this case, we can solve {\it halting} of the Game of Life simply by simulating the dynamics for (possibly exponential in the size of a graph) number of iterations.\\

In this section, we study the halting problem for $GLG(G, A_0, 1, 1, 1)$. We select this particular instance in order to be consistent with the following section on Graph Isomorphism testing. We illustrate that the complexity of the Game of Life follows the phase transition behavior with respect to graph density.\\

Studying phase transitions is a rich topic at the intersection of computer science and statistical physics. Random instances of many different problems and families of structures in computer science undergo a phase transition with respect to one or more parameters. Moreover, the most challenging and complex instances are observed around the transition point, while the instances far from the transition are usually simpler \cite{kSATPhase, NetworkPhase, biamonte}.\\

One of the first famous results on this topic was about 3-SAT phase transition \cite{3SAT, 3SATPhase}. In that seminal paper, the authors consider random instances of the boolean satisfability problem with $N$ variables, $M$ clauses, and $3$ variables per clause. For large enough values of $N$ (what is called thermodynamic limit in physics) the random instances undergo a sharp SAT-UNSAT phase transition with respect to density parameter $k = \frac{M}{N}$ at some value $k=\alpha$: for $k < \alpha-\epsilon$ random instances are satisfiable with probability one while for $k > \alpha+\epsilon$ random instances are unsatisfiable with probability one.\\

Another well-known example is the phase transition in Erdos-Renyi random graphs \cite{ERPhase}. Studying the phase transitions of computer science problems is a rich and ongoing line of research; see for example \cite{CSP} and references therein. 

\begin{figure}[ht]
\centering
\begin{subfigure}{.45\textwidth}
    \centering
    \includegraphics[width=0.9\columnwidth]{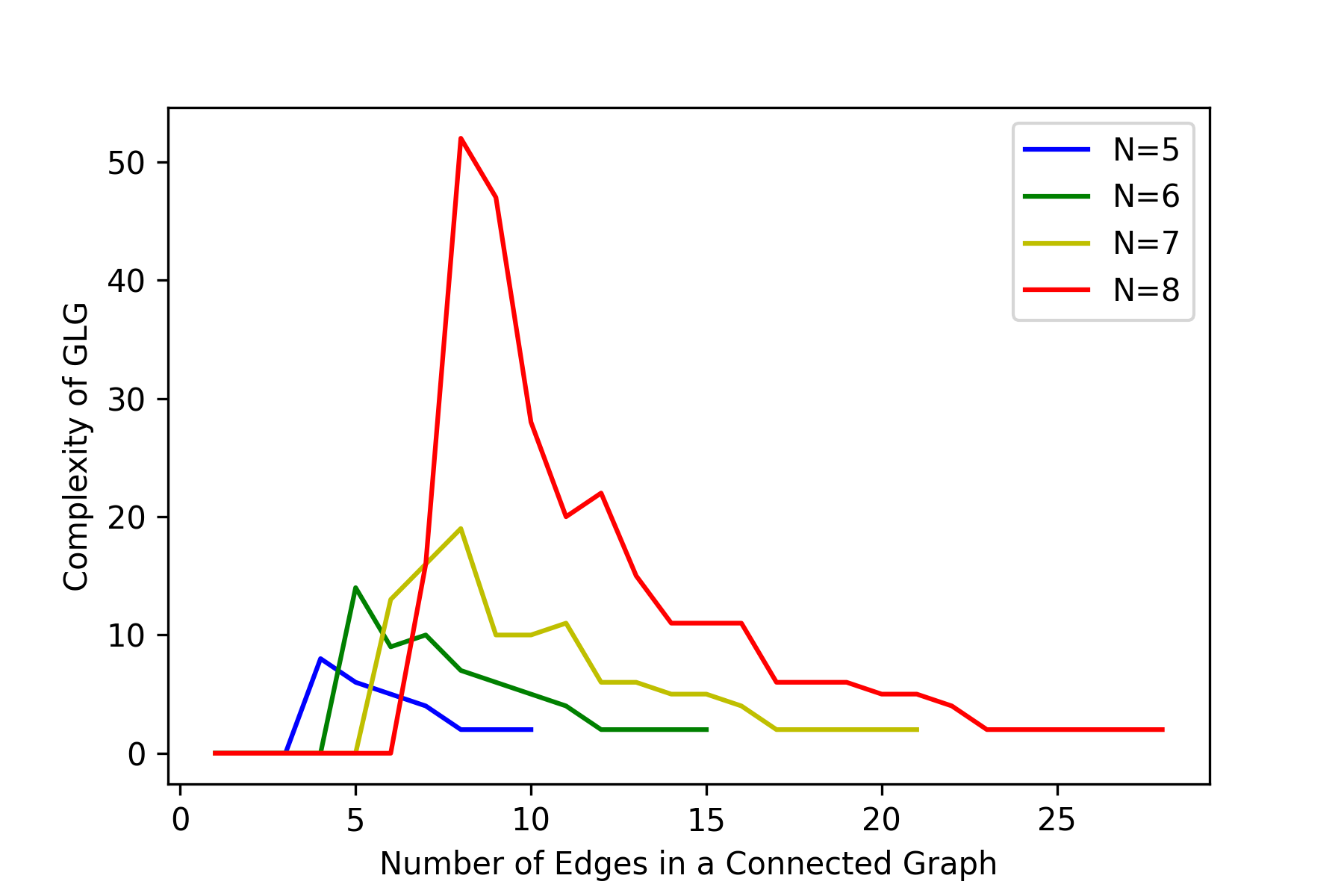} 
    \caption{Complexity of $GLG$ for the set of all small connected graphs with up to 8 vertices.} 
    \label{fig:connected_complexity}
\end{subfigure}%
\hfill
\begin{subfigure}{.45\textwidth}
    \centering
    \includegraphics[width=0.9\columnwidth]{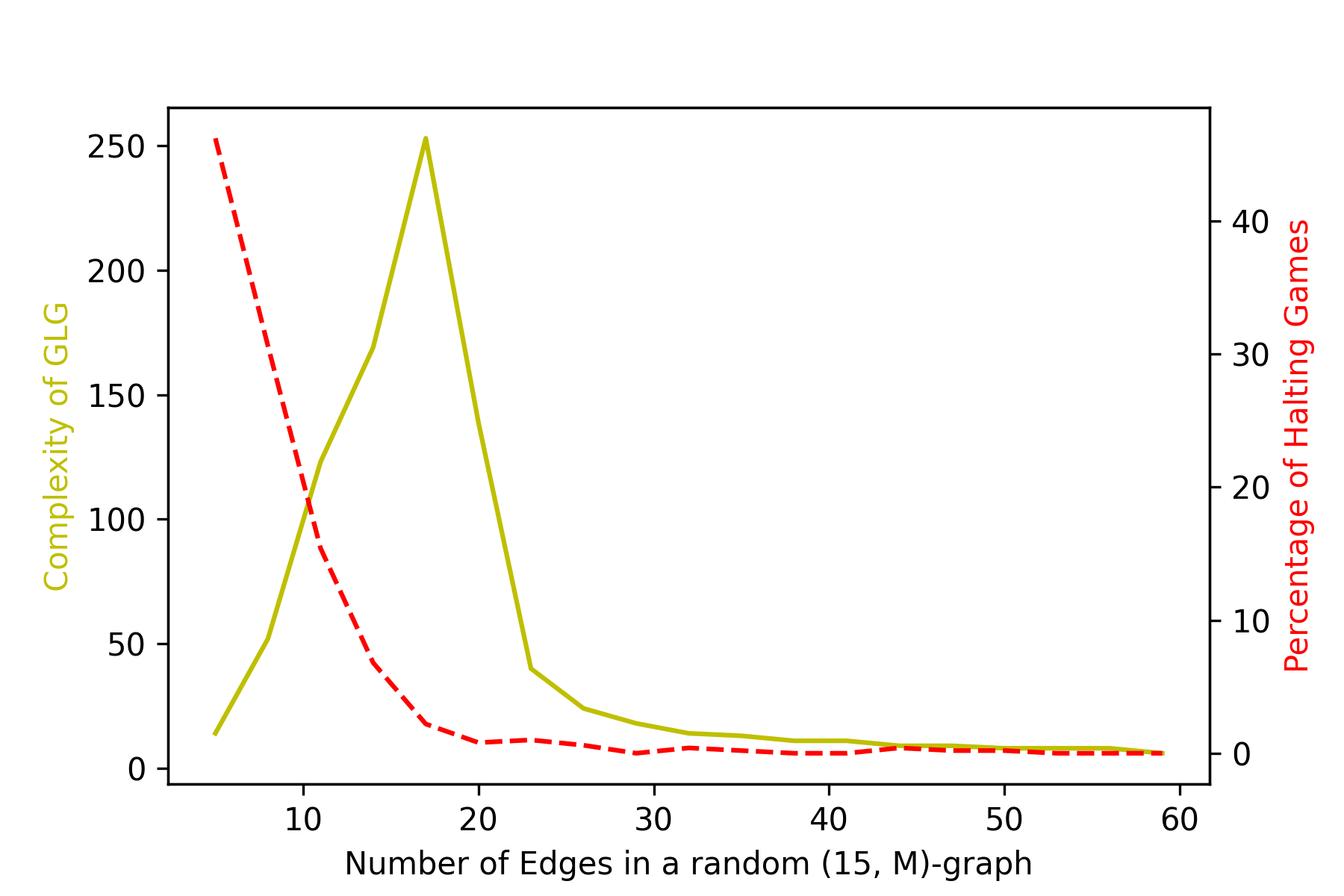}
    \caption{Complexity of $GLG$ and the percentage of halting games for small $G(N, M)$-random graphs.}
    \label{fig:random_complexity}
\end{subfigure}
\caption{Complexity of the Game of Life on Graphs.}
\label{fig:complexity_both}
\end{figure}

In Figure~\ref{fig:complexity_both} we illustrate the similar behavior of the Game of Life on Graphs. In the Figure~\ref{fig:connected_complexity} we consider the set of all connected graphs with $N$ vertices for $N\in \{5,6,7,8\}$. For every such graph, we simulate $N$ instances of GLG (one for every vertex serving as initially 'alive'). For every GLG we compute its complexity (the number of non-repeating {\it life patterns}). We observe that the most complex patterns appear at a particular edge density ($M \sim cN$).\\

In the Figure~\ref{fig:random_complexity} we consider random $G(N,M)$-graphs. For $N=15$ and every $M \le 60$ we sample 500 random graphs and simulate $N$ instances of GLG (one for every vertex serving as initially 'alive'). We compute the number of non-repeating {\it life patterns} and the percentage of {\it halting} games. We observe the behavior that resembles ensembles of random NP-complete problems and conjecture that the most difficult instances for {\it halting} problem appear at the particular density ($M \sim cN$) while for other densities {\it halting} can be solved simply simulating the dynamics.

\section{Isomorphism Testing}
\label{sec:GI}

{\bf Definition.} Two graphs $G = (V_G, E_G)$ and $H = (V_H, E_H)$ are called {\it isomorphic} iff there exists a bijection $f: V_G \to V_H$, such that $(a, b) \in E_G \Leftrightarrow (f(a), f(b)) \in E_H$ for every $a,b \in V_G$. Speaking informally, graphs $G$ and $H$ are isomporhic, if there exists a bijection of vertices that preserves incidence.\\

The Graph Isomorphism problem is not known to be equivalent to any of NP-complete problems. At the same time, no polynomial algorithm for testing graph isomorphism is known; the best-known algorithm runs in quasipolynomial time \cite{babai}.  Thus the Graph Isomorphism is known as the problem of intermediate complexity. We refer to \cite{intro, thesis} as excellent introductions to graph isomorphism and known heuristics.\\

One of the goals of this paper is to introduce another powerful approach to testing graphs isomorphism. We conjecture that the highly irregular structure of the patterns produced by the Game of Life on Graphs can be used for efficient graph isomorphism testing.

\subsection{Feature extraction}

There are multiple ways to extract features from the Game of Life on Graphs. In this work, we describe the most straightforward way that is similar to the famous WL algorithm. The approach described in this section is simpler since we consider integer labels instead of multisets and hashes. Meanwhile, our approach generalizes WL algorithms in the following sense: we update labels on vertices with respect to complex {\it life patterns} while WL algorithm update depends only on neighbors of a vertex.\\

Consider a graph $G = (V, E)$, fix the iteration parameter $k \ge 1$ and do the following:
\begin{enumerate}[leftmargin=*]
    \item Put a real label $l^i_0 = 1.0$ at every vertex $i \in V$.
    \item Fix $A_0^i = \{i\}$ for every $i \in V$.
    \item Initialize $n = |V|$ instances of the Game of Life on Graphs: \[GLG(G, A_0^i, 1, 1, 1)\]
    \item Compute $k$ steps of evolution for each of $n$ instances of $GLG$. \item Update labels step by step for $t \in [1, ..., k]$ iterations and simultaneously for all vertices: \[l^i_t = l^i_{t-1} + \sum\limits_{v\in A_{t}^i} l^v_{t-1}\]
    \item (Optionally) Normalize all the the labels at step $t$: \[l^i_t = l^i_t / \left (\sum\limits_{j\in V} l^j_{t} \right )\]
    \item For every $t \in [1, ..., k]$ for a vector $f_t = (l^0_t, ..., l^n_t)$. Sort it in the increasing order.
    \item The resulting vector of features is the concatencation of vectors $f_t$ for every $t \in [1, ..., k]$.
\end{enumerate}

{\bf Proposition.} If two graphs are isomorphic, all its life patterns $A_t^i$ coincide under the isomorphism, so are their label sets $l^i_t$ and the resulting feature vectors.\\

We describe an isomorphism testing procedure in Algorithm~1. Here we follow our feature extraction algorithm step by step for each of the two graphs. If at a step $t$ sorted label sets do not coincide, we have the non-isomorphism certificate. If all the label sets coincide, the graphs are likely to be isomorphic.\\

\begin{center}
    \begin{algorithm}[H]
 \KwData{Graphs $G = (V_G, E_G)$ and $H = (V_H, E_H)$, number of iterations $k$.}
 \KwResult{{\bf True}, if graphs $G$ and $H$ are likely to be isomorphic;\newline {\bf False}, if graphs $G$ and $H$ are provably non-isomorphic.}
 $t=1$\;
 \For{$v_G\in V_G$} {
 $l_0^{v_G}=1.0$, $A_0^{v_G}=\{v_G\}$, $GLG(G, A_0^{v_G}, 1, 1, 1)$\;
 }
 \For{$v_H\in V_H$} {
 $l_0^{v_H}=1.0$, $A_0^{v_H}=\{v_H\}$, $GLG(H, A_0^{v_H}, 1, 1, 1)$\;
 }

 \For{$t \in [1,...,k]$}{
  Compute $A_t^{v_G}$ for $v_G\in V_G$\;
  Compute $A_t^{v_H}$ for $v_H\in V_H$\;
  Update $l_t^{v_G}$ for $v_G\in V_G$\;
  Update $l_t^{v_H}$ for $v_H\in V_G$\;
  $g_t$ = {\bf sort} $(l_t^{v_G})_{v_G\in V_G}$\;
  $h_t$ = {\bf sort} $(l_t^{v_H})_{v_H\in V_H}$\;
  \If{$g_t \ne h_t$}{
   Return {\bf False}\;
   }
 }
 Return {\bf True}\;
 \caption{Graph Isomorphism Testing.}
\end{algorithm}
\end{center}

There are various possible generalizations of the feature extraction algorithm above; we list a couple of them:
\begin{enumerate}[leftmargin=*]
    \item consider multiset labels $l_t^i = A_t^i$. Use this sets to get a canonical form or simply calculate co-occurence statistics.
    \item Consider $l_t^i$ as hashes. At step 5) of the feature extraction algorithm, sort hashes $l_{t-1}^v$ and concatenate them.
\end{enumerate}


\section{Experiments}
\label{sec:exp}

We have tested our algorithms on the graph collections from \url{http://users.cecs.anu.edu.au/~bdm/data/graphs.html}. We observe that for every pair of non-isomorphic connected graphs with up to ten vertices, Algorithm~1 is able to provide a non-isomorphism certificate in only two steps. The same is true for small connected regular graphs. Our code (python within jupyter notebook) is available at \url{https://github.com/mkrechetov/GameOfLifeOnGraphs}.\\

Another (trivial) observation was that the euclidean distance induced by our feature vectors satisfies triangle inequality for all small connected graphs with up to ten vertices. To measure the distance between two graphs, we compute their feature vectors $f_t$ described in the previous section. Then we calculate the euclidean distance between that two vectors. We call it 'GLG'-distance between two graphs and denote it $d_{GLG}(G, H)$.\\

\begin{figure}[ht]
\centering
\includegraphics[width=0.33\textwidth]{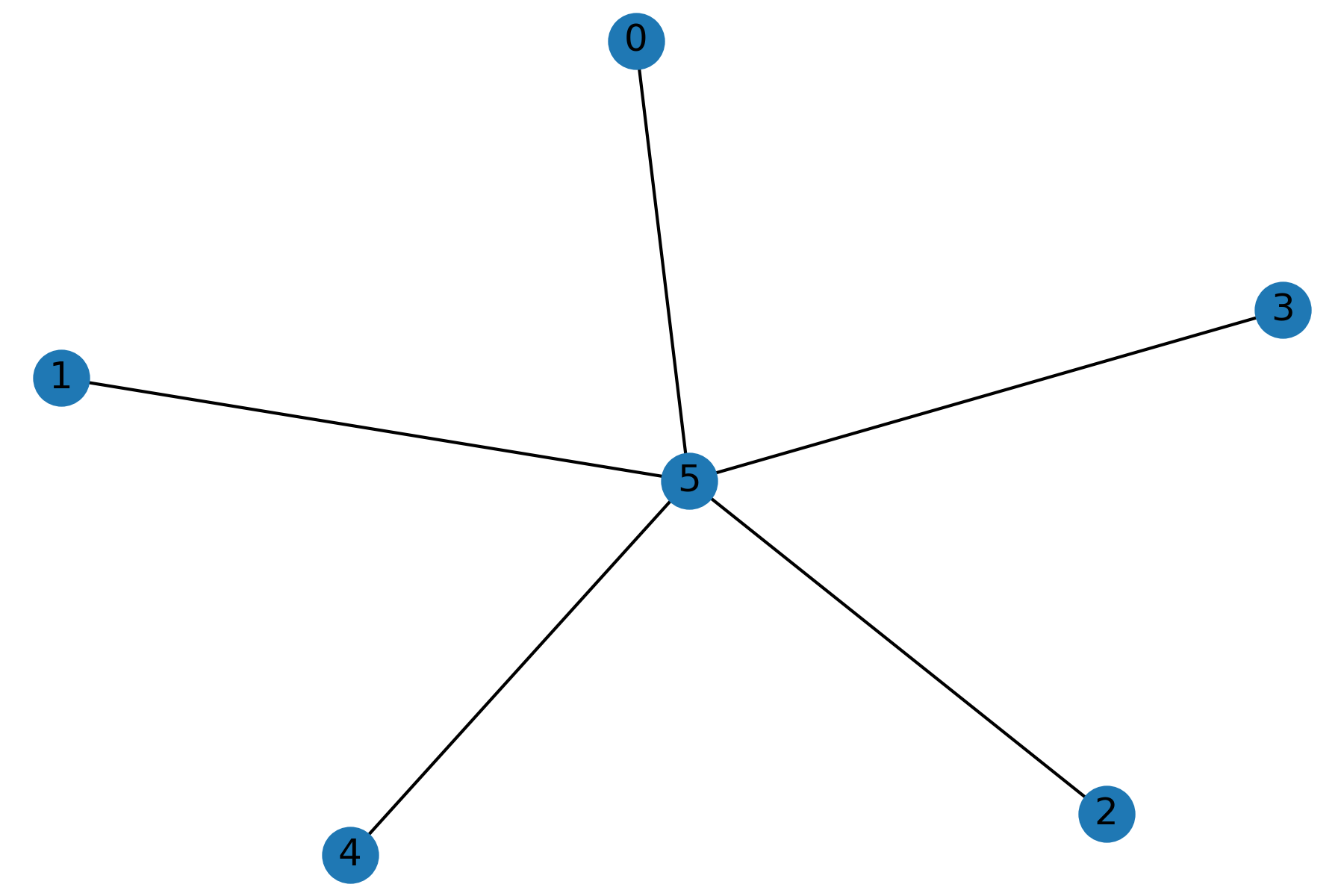}\hfill
\includegraphics[width=0.33\textwidth]{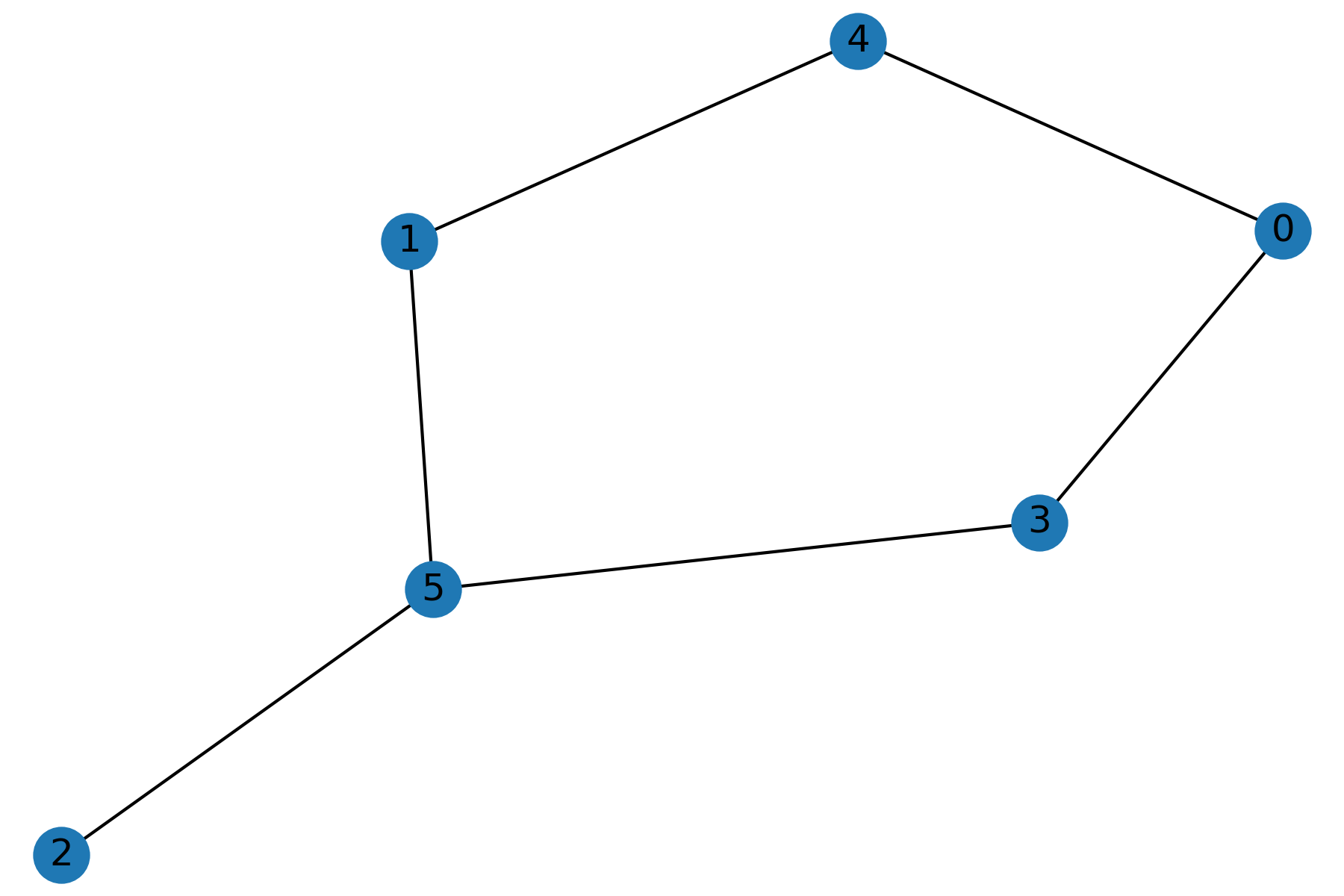}\hfill
\includegraphics[width=0.33\textwidth]{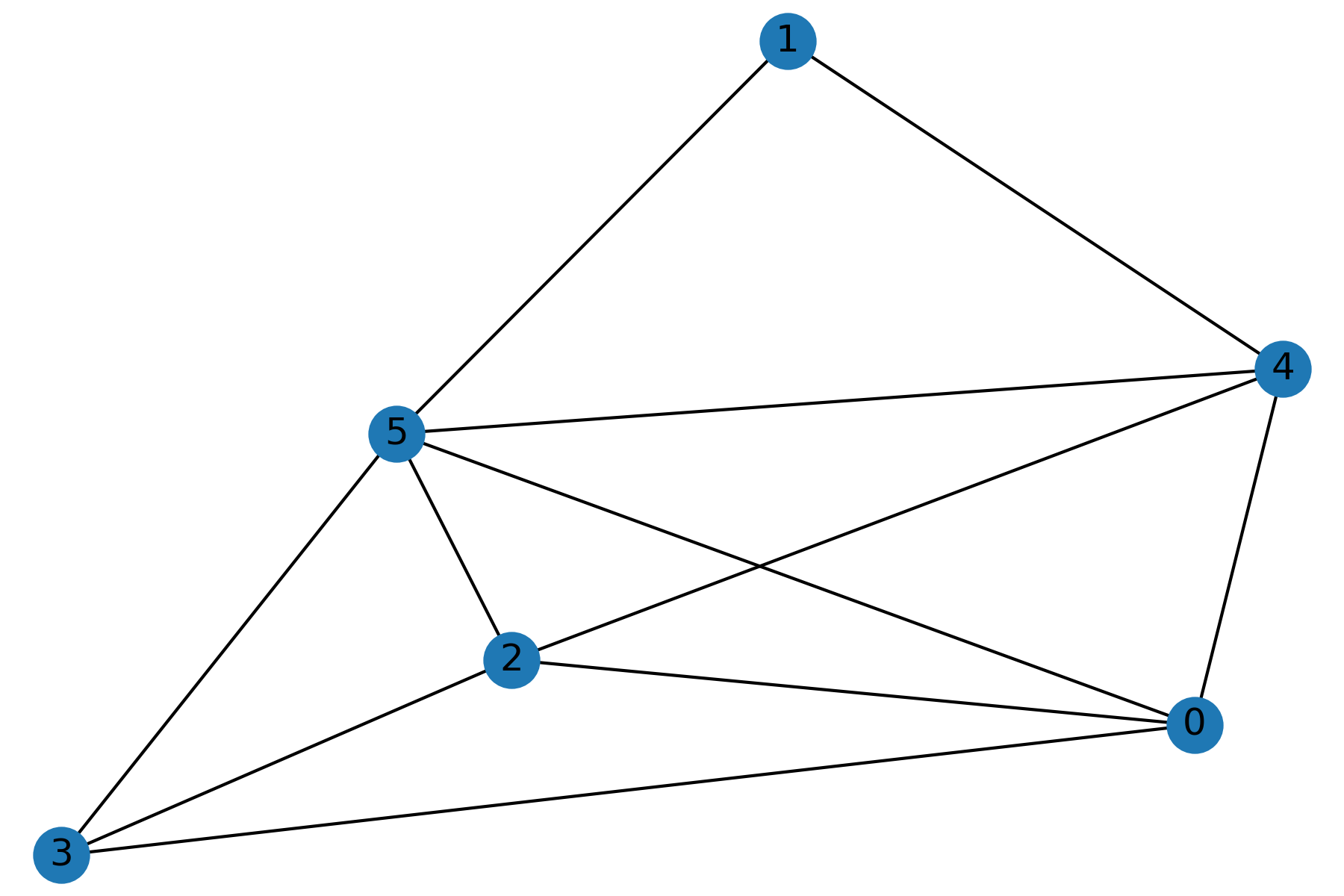}\\
\includegraphics[width=0.33\textwidth]{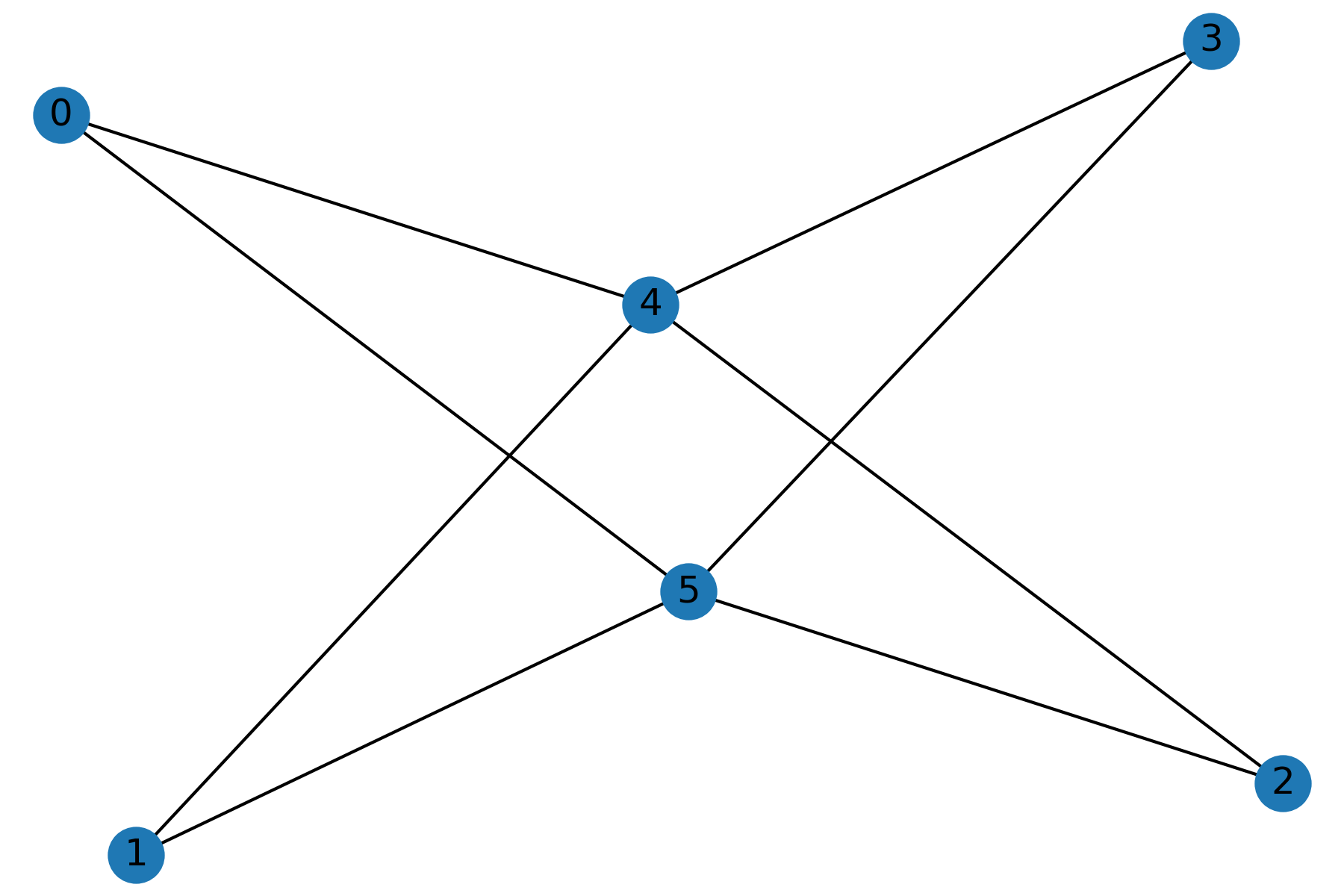}\hfill
\includegraphics[width=0.33\textwidth]{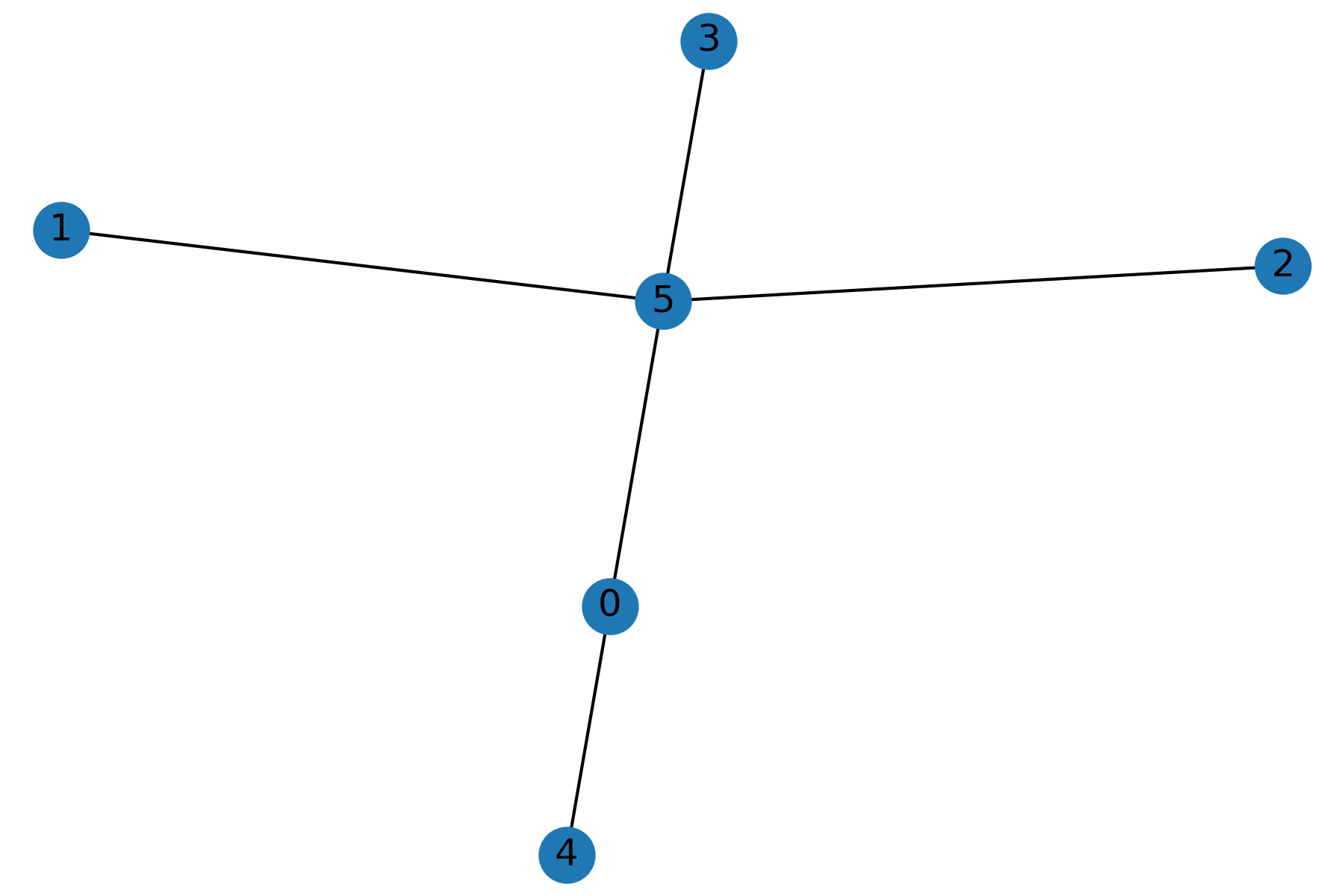}\hfill
\includegraphics[width=0.33\textwidth]{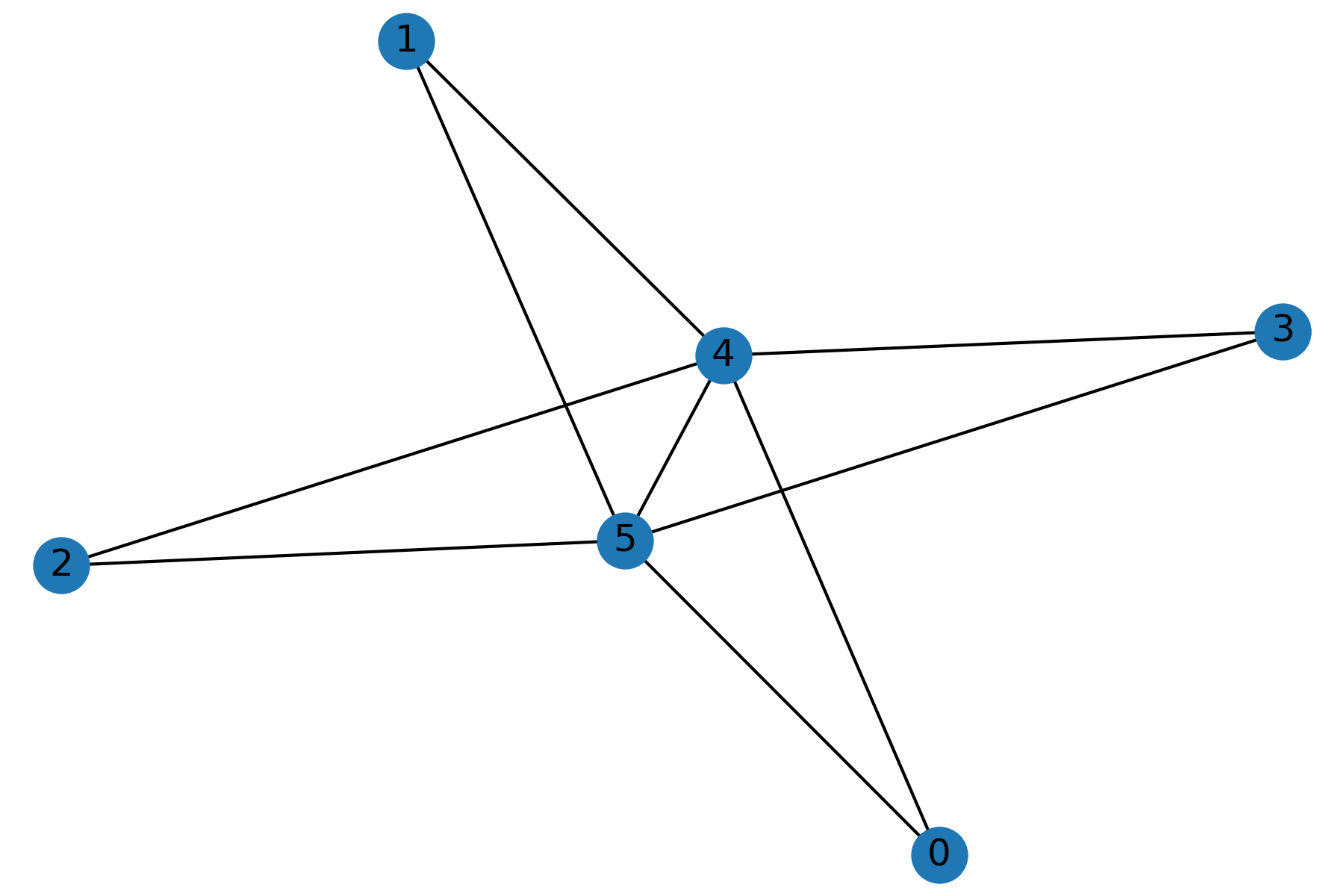}\\
\includegraphics[width=0.33\textwidth]{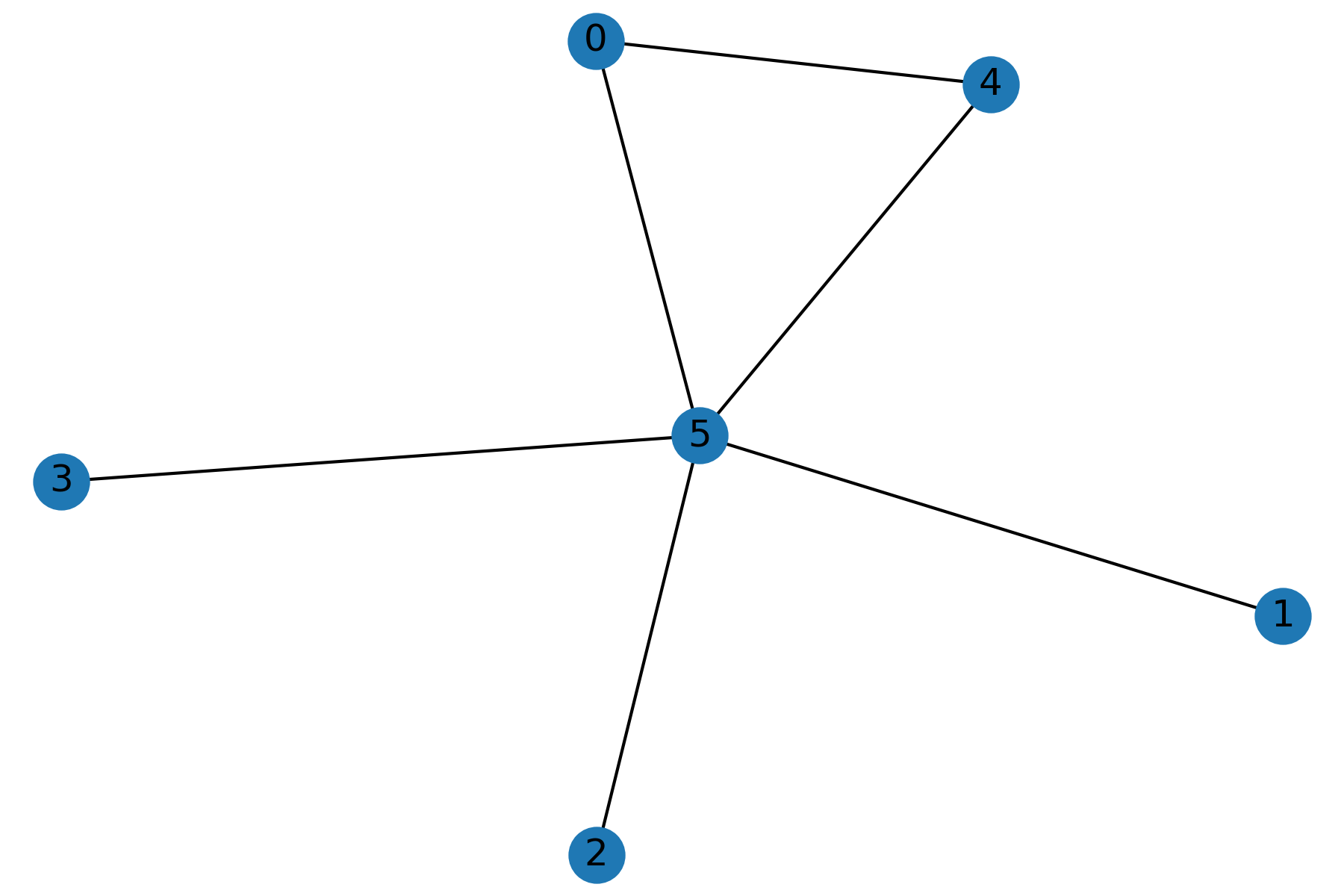}\hfill
\includegraphics[width=0.33\textwidth]{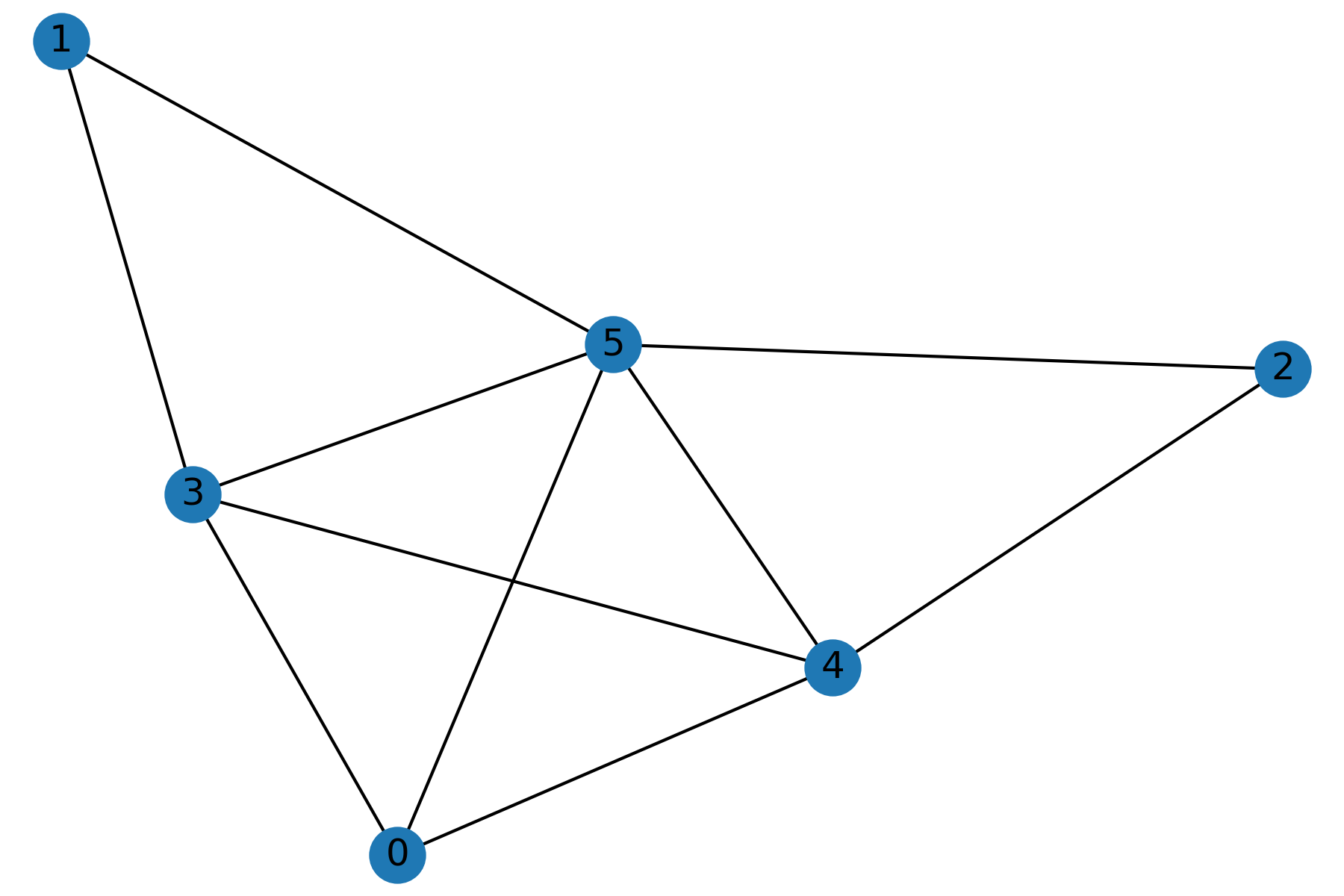}\hfill
\includegraphics[width=0.33\textwidth]{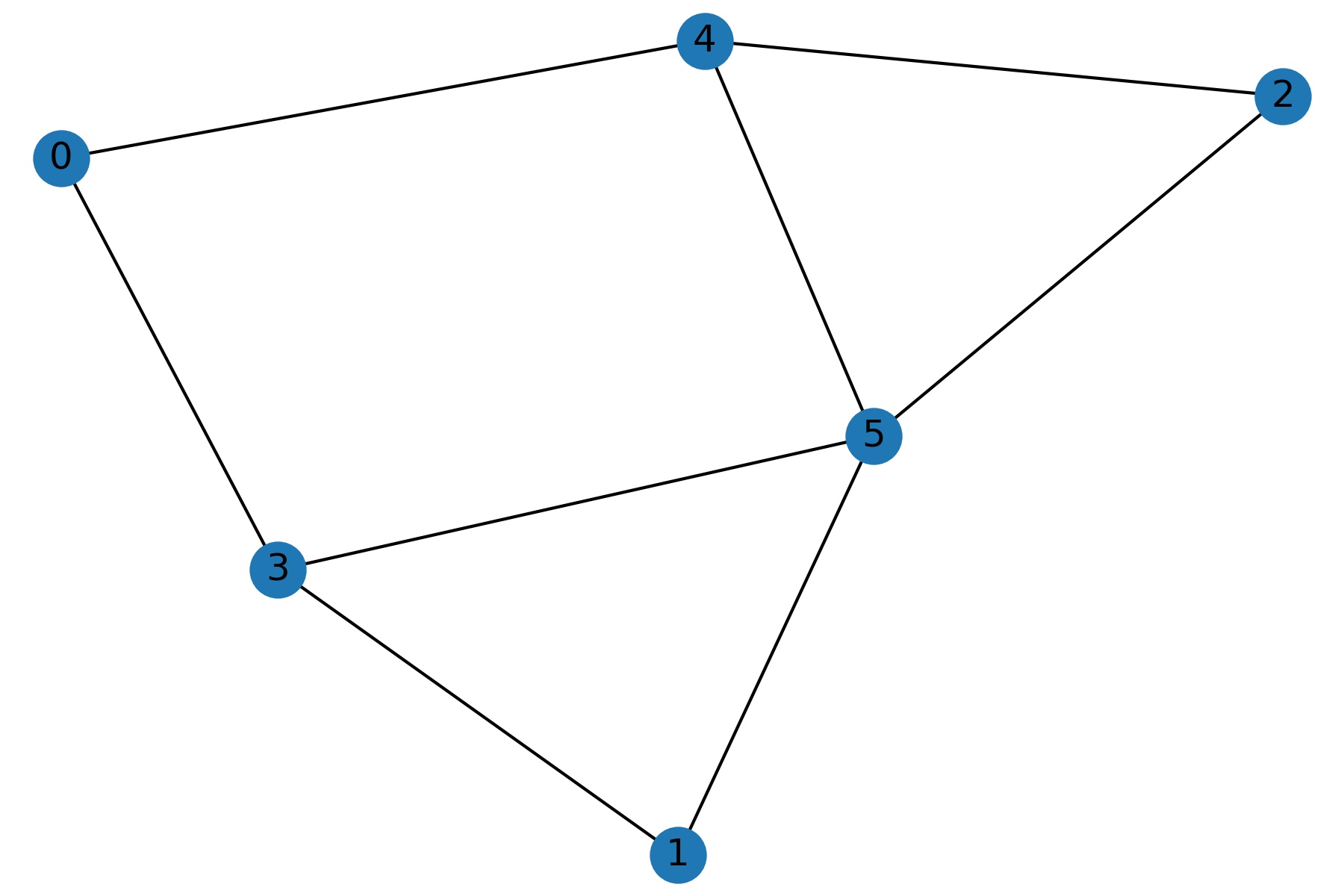}\\
\includegraphics[width=0.33\textwidth]{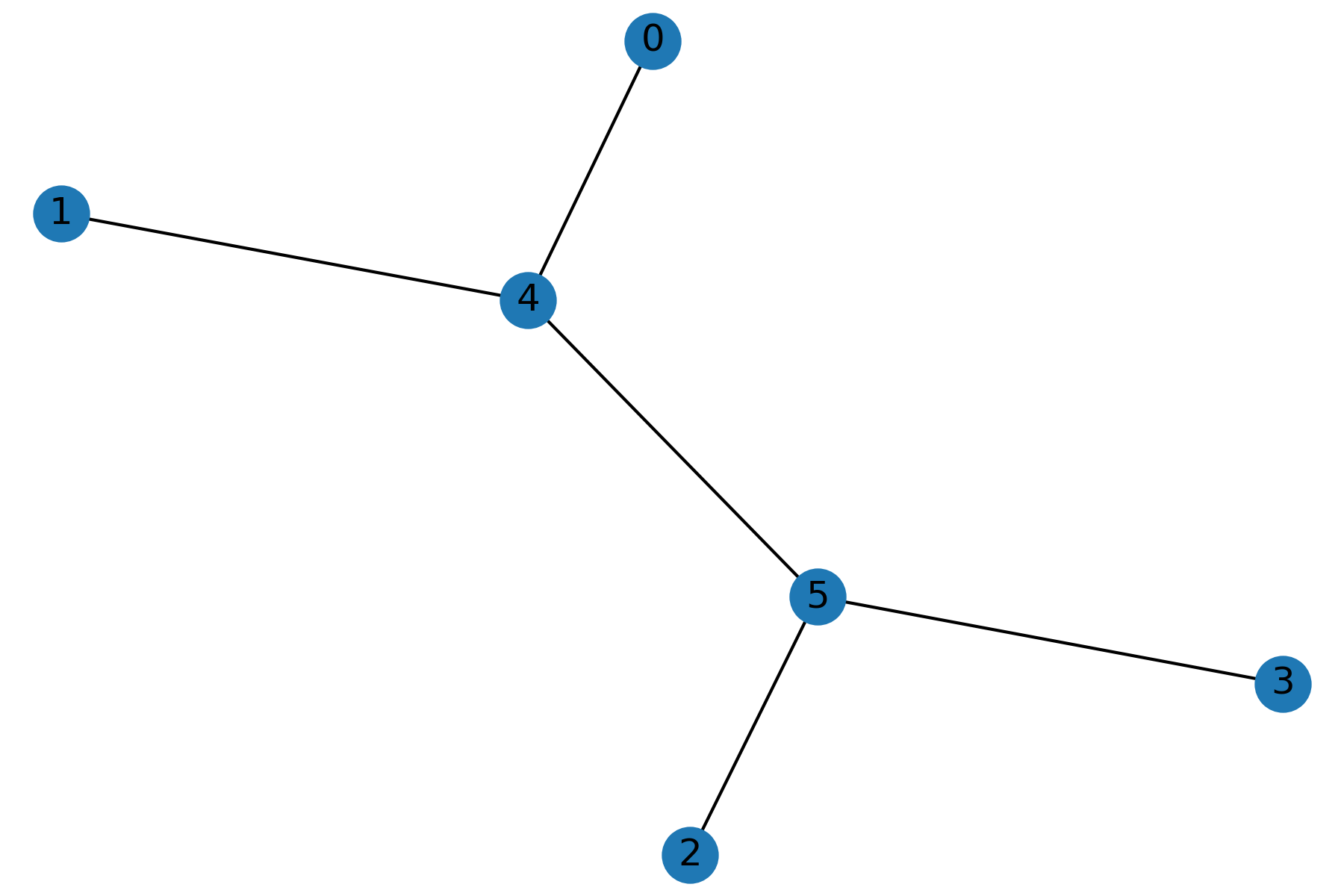}\hfill
\includegraphics[width=0.33\textwidth]{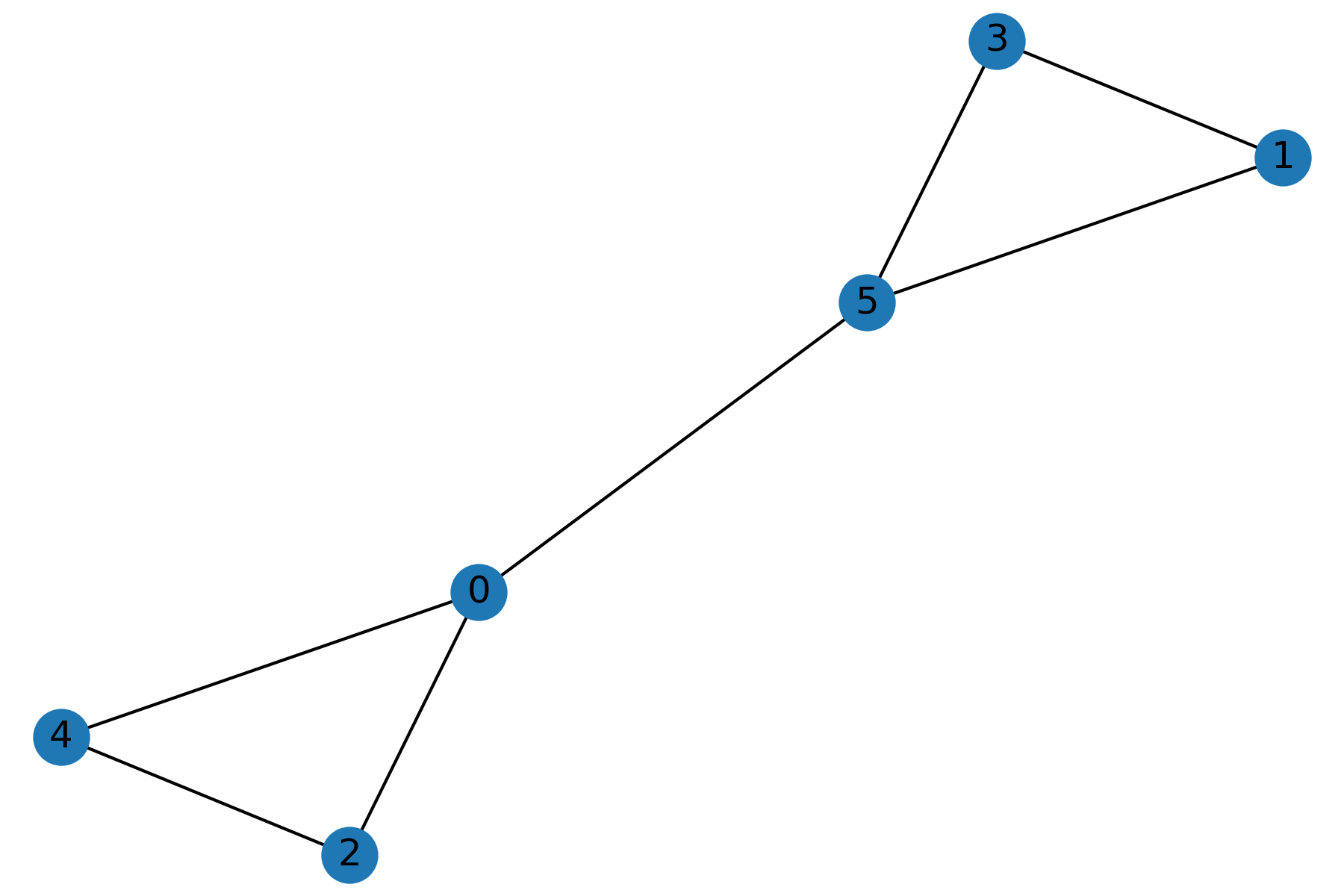}\hfill
\includegraphics[width=0.33\textwidth]{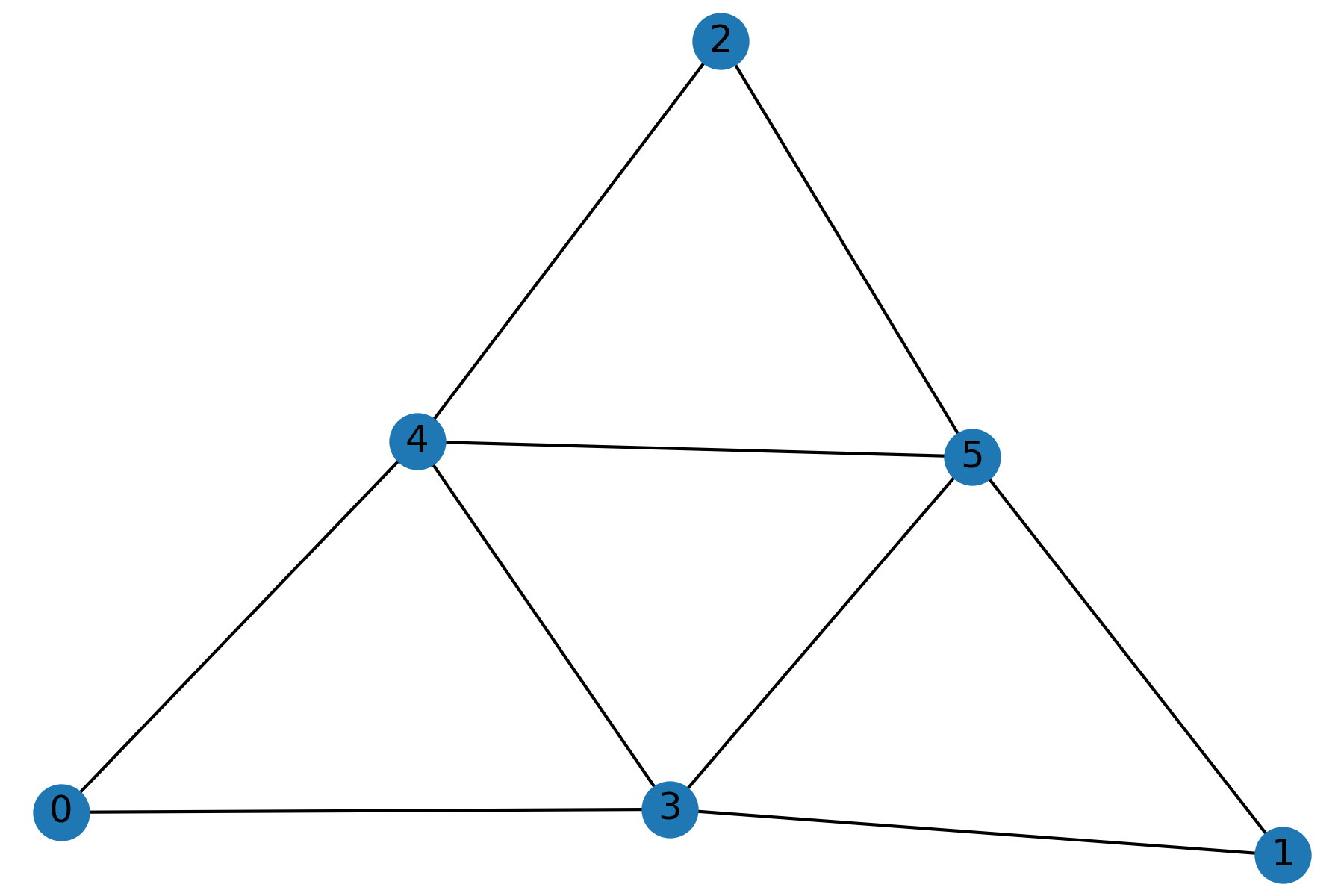}\\
\includegraphics[width=0.33\textwidth]{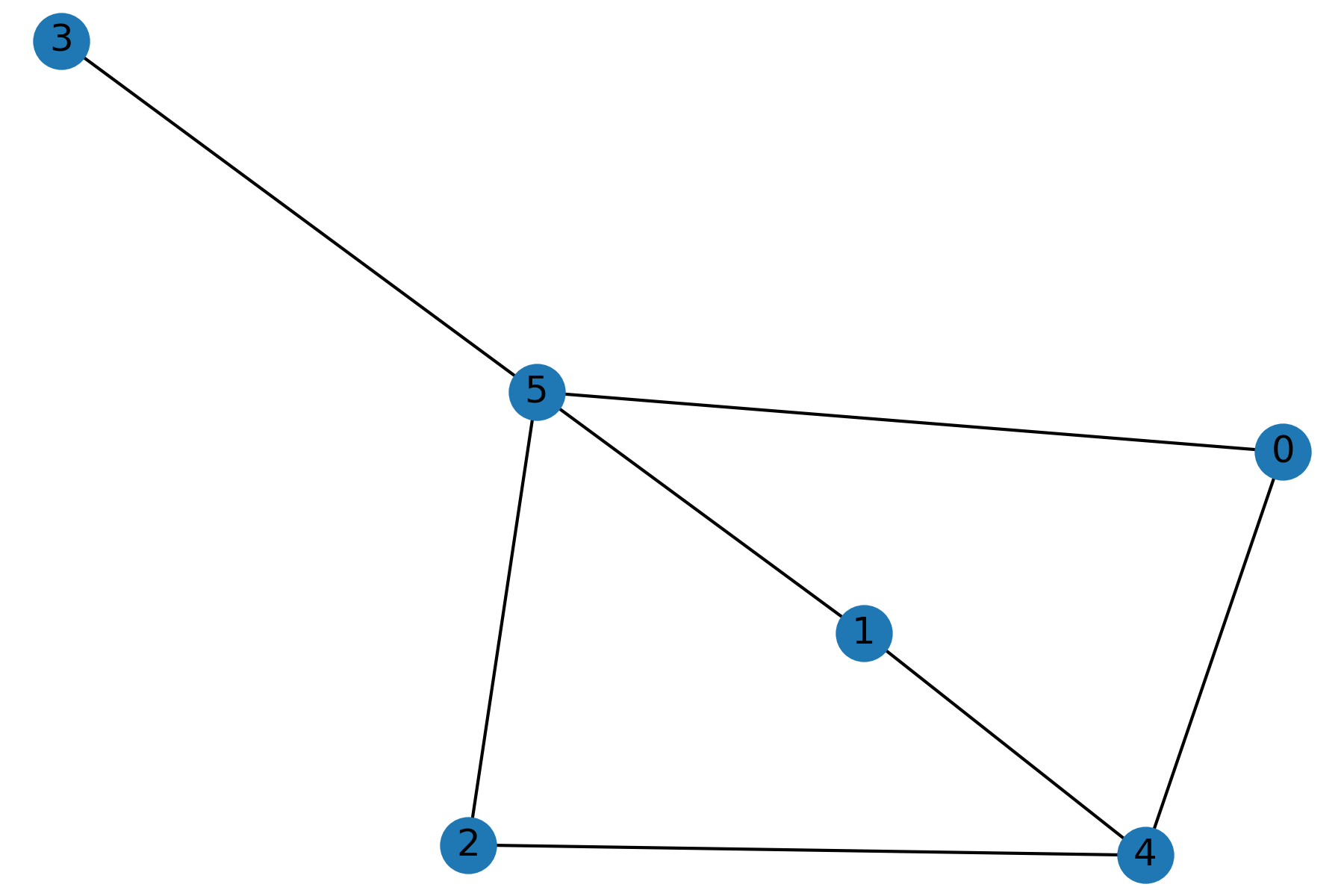}\hfill
\includegraphics[width=0.33\textwidth]{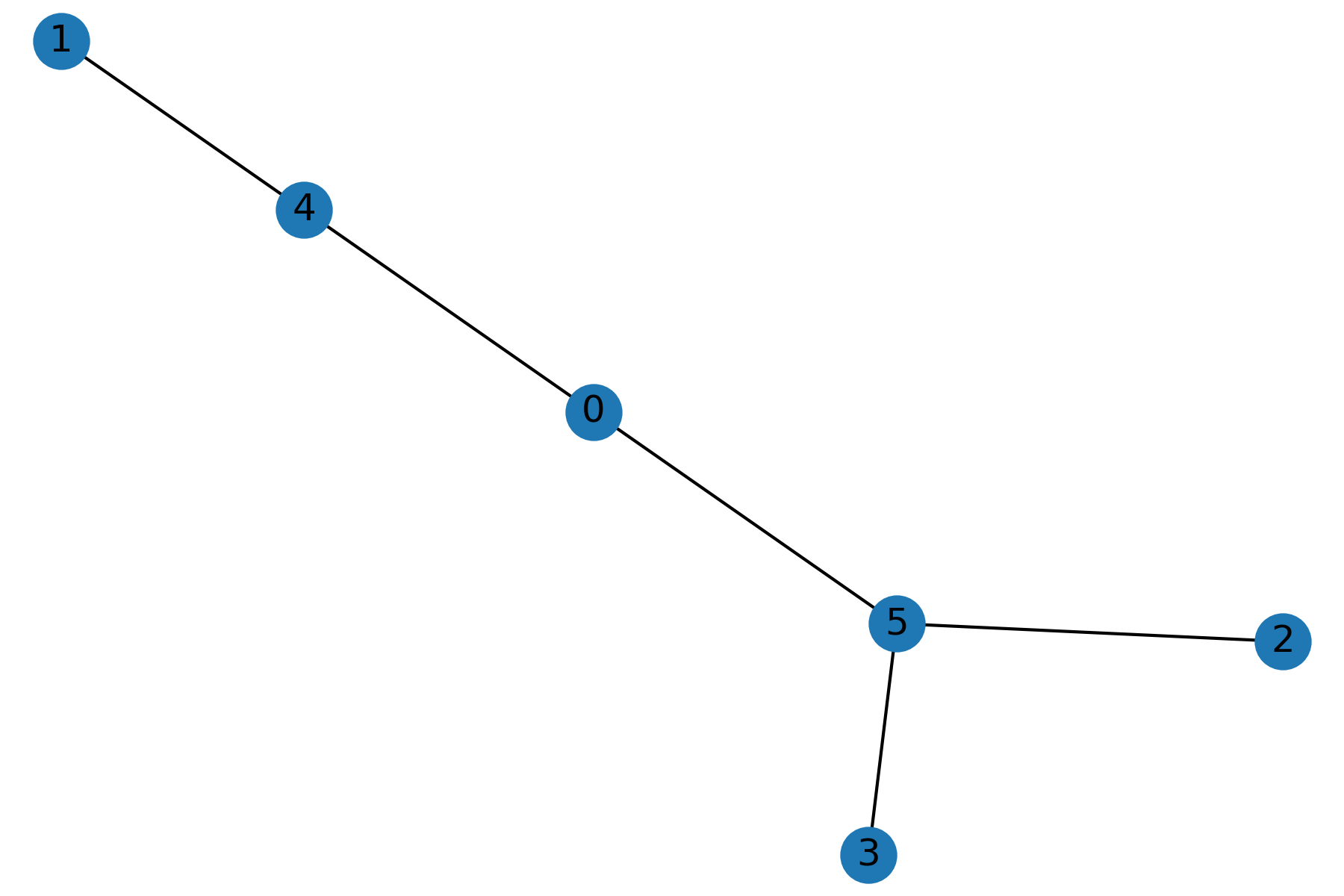}\hfill
\includegraphics[width=0.33\textwidth]{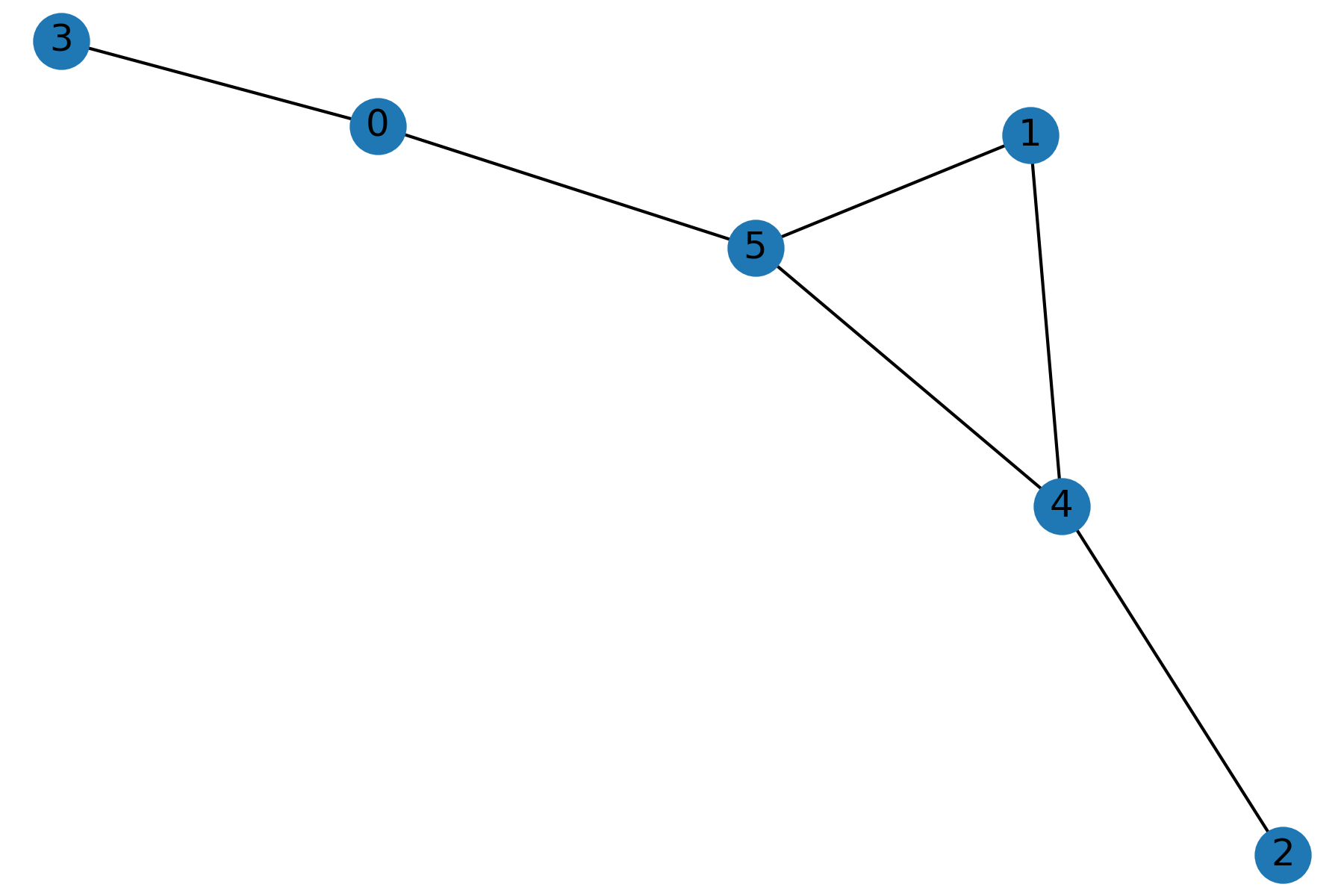}\\
\caption{Some examples of lines in the space of all graphs with six vertices. For every row, the 'GLG'-distance between the first and thrid graphs equals the sum of the two other distances. The graph in the middle can be interpreted as weighted sum of two other graphs in a row.}
\label{fig:metric_lines}
\end{figure}

We observe that the triangle inequality becomes equality for some graphs. If a graph triple $(G_1, G_2, G_3)$ satisfies the following equality:
\[
d_{GLG}(G_1, G_2) = d_{GLG}(G_1, G_3) + d_{GLG}(G_2, G_3),
\]
we may interpret the graph $G_3$ as a weighted sum of graphs $G_1$ and $G_2$. We list some examples in Figure~\ref{fig:metric_lines} and conjecture that this observation may lead to some future topological interpretations.

\section{Conclusions and Path Forward}

In this work, we defined the discrete dynamics we call Game of Life on Graphs. We demonstrated how the rich structure of the life patterns could be used for graph isomorphism testing. We also showed that the Game of Life might be used to define metric space on the set of small connected graphs with up to 10 vertices. Hopefully, the concept of the Game of Life on Graphs will find more applications in Graph Theory and Computer Science.\\

We conclude the manuscript with an incomplete list of further research directions, presented in the order of importance (subjective): 

\begin{itemize}[leftmargin=*]
    \item Are there examples of non-isomorphic graphs that are not distinguished by one or another version of the Game of Life on Graphs? If they exist, then specify the family of graphs for which the Graph Isomorphism problem is solved by the Game of Life on Graphs.
    
    \item What is the computational complexity of {\it halting} of the Game of Life on Graphs for different values of $\mathfrak a, \mathfrak d, \mathfrak r$? Does the Complexity of GLG undergo the sharp phase transition in the thermodynamic limit $N\to \infty$? If so, what is the value of edge density $p(N)$ at the transition point? 
    
    \item Can we use the Game of Life to extract features with good predictive power for graph classification problems? We conjecture that the features extracted from the Game of Life on Graphs are at least as descriptive as the features by DeepWalk \cite{DeepWalk} and WL kernels \cite{WLKernels}. Moreover, the features extracted from the Game of Life are completely deterministic, which can be beneficial for machine learning applications.
    
    \item Does the metric induced by the Game of Life on small graphs capture topological properties of graphs? Can the operation of the weighted sum from the Figure~\ref{fig:metric_lines} be described in a simpler way, e.g., as an operation on edges/cycles of corresponding graphs?
\end{itemize}

\bibliography{biblio}
\bibliographystyle{plain}

\end{document}